\documentstyle[twoside]{article}
\begin{document}
%
%
\begin{center}
{\large\bf 
Some Aspects of
Generalized Contact Interaction in
One-Dimensional Quantum Mechanics}
\end{center}
\begin{center}
Taksu Cheon$^{1,2}$ and T. Shigehara$^{3}$ \\
{\small
{\it
$^{1}$ Laboratory of Physics, Kochi University of Technology \\
Tosa Yamada, Kochi 782-8502, Japan\\
$^{2}$ Theory Division, 
High Energy Accelerator Research Organization (KEK) \\
Tsukuba, Ibaraki 305-0801, Japan \\
$^{3}$ Department of Information and Computer Sciences\\
Saitama University, 
Urawa, Saitama 338-8570, Japan \\
}
}
\end{center}
\abstract
We construct a one-dimensional contact interaction 
($\varepsilon$ potential) which induces the discontinuity of 
the wave function while keeping its derivative continuous.
By combining the $\varepsilon$ potential and the Dirac's 
$\delta$ function, 
we construct most general one-dimensional contact interactions 
allowable under the time reversal symmetry. 
We present some elementary results for the scattering problem 
which suggest a dual relation between $\delta$ and 
$\varepsilon$ potentials.
\endabstract

\bigskip
\bigskip
%
%
\noindent{\large\bf I. Introduction}

\medskip
The quantum-mechanical
contact interaction is a natural idealization of an obstacle which
is much smaller than the wavelength of a particle [1]. 
In spite of the seeming simplicity of contact interactions, 
there are several non-trivial aspects which are largely
left unexplored. 
Even in the simplest setting of one dimension,
there has been a longstanding problem of approximating 
the generalized contact interaction 
with a short-range limit of a local self-adjoint operator. 
Apart from a rather trivial case of Dirac's $\delta$ potential, 
generalized contact interaction has resisted the realization
of such representation.  Existing approximations have been
either non-local or non-Hermitian [2-6], thus have left us 
with little physical intuition.  Also, one could not find much
relevance to the
experimentally realizable ``real-life'' quantum mechanics.
In this paper, we sketch a scheme for just such 
local realization of generalized contact interaction.  We also
show that the object so realized
is not only a natural generalization, but also 
an indispensable element
of one-dimensional quantum mechanics.

We consider the one-dimensional Schr\"{o}dinger equation 
\begin{eqnarray}
\label{eq1:1}
-\varphi''(x) + V(x) \varphi(x) = k^2 \varphi(x), 
\end{eqnarray}
where $V(x)$ belongs to  
a general class of quantum-mechanical contact interactions; 
$V(x)=0$ for $x\neq 0$. 
The self-adjoint extension theory dictates [1] that 
if the system has the time reversal symmetry, 
there exists  a three parameter family of solutions, 
characterized by the connection conditions
\begin{eqnarray}
\label{eq1:2}
\left( \! \!
\begin{array}{c}
\varphi'(0_+) \\
\varphi (0_+) 
\end{array}
\! \! \right)
=
{\cal V}
\left( \! \!
\begin{array}{c}
\varphi'(0_-) \\
\varphi (0_-) 
\end{array}
\! \! \right) 
\end{eqnarray}
with ${\cal V} \in SL(2,{\bf R})$. 
Clearly, the connection matrix 
\begin{eqnarray}
\label{eq1:3}
{\cal V}_{\delta}(v) = 
\left(
\begin{array}{cc}
1 & v \\
0 & 1 
\end{array}
\right)
\end{eqnarray}
corresponds to the $\delta$ potential of strength $v$;
\begin{eqnarray}
\label{eq1:4}
V(x)=v\delta(x). 
\end{eqnarray}
Our main purpose is
to show that a potential consisting of several
$\delta$ functions with disappearing distances 
can give rise to all possible connection conditions Eq.(\ref{eq1:2}) 
with appropriate scaling of the coupling strengths.
This means that all allowable one-dimensional 
contact interactions can be realized 
as the zero-range limit of local self-adjoint (namely, 
physically realizable) interactions. 
We start in Sect.II by constructing a zero-range force 
which leaves
the derivative of the wave function $\varphi'(x)$  
continuous, but makes  $\varphi(x)$ itself  discontinuous [2,3]; 
\begin{eqnarray}
\label{eq1:5}
{\cal V}_{\varepsilon}(u) = 
\left(
\begin{array}{cc}
1 & 0  \\
u & 1 
\end{array}
\right). 
\end{eqnarray}
We denote the contact interaction with the connection condition 
(\ref{eq1:5}) by $\varepsilon(x)$;  
\begin{eqnarray}
\label{eq1:6}
V(x)=u\varepsilon(x). 
\end{eqnarray}
We then show that the general connection condition of the form 
Eq.(\ref{eq1:2}) can be constructed from 
the $\varepsilon$ and $\delta$ potentials. 
Some elementary results for the scattering problem are presented 
in Sect.III, where a dual role of $\delta$ and $\varepsilon$ potentials 
is emphasized. This work is summarized in Sect.IV. 

\bigskip
\bigskip
%
%
\noindent{\large\bf II.
Local Realization of Discontinuity-Inducing \\
Contact Interactions} 

\medskip
We begin with the Schr\"{o}dinger equation 
in one-dimensional free space; 
\begin{eqnarray}
\label{eq2:1}
-\varphi''(x) = k^2 \varphi(x), 
\end{eqnarray}
where $k$ is the wave number of a particle. 
It is convenient to introduce a vector notation 
for the wave function and its space derivative;   
\begin{eqnarray}
\label{eq2:2}
{\bf \Psi}(x)=
\left( \! \!
\begin{array}{c}
\varphi'(x) \\
\varphi(x) 
\end{array}
\! \! \right) .
\end{eqnarray}
The Schr\"{o}dinger equation, Eq.(\ref{eq2:1}) now takes the form
\begin{eqnarray}
\label{eq2:3}
{\bf \Psi}'(x) = {\cal H}(k)
{\bf \Psi}(x)  
\end{eqnarray}
with   
\begin{eqnarray}
\label{eq2:4}
{\cal H}(k) = 
\left(
\begin{array}{cc}
0 & -k^2  \\
1 & 0
\end{array}
\right). 
\end{eqnarray}
The solution of Eq.(\ref{eq2:3}) is given by 
\begin{eqnarray}
\label{eq2:5}
{\bf \Psi}(x) = {\cal G}(k;x-x_0) {\bf \Psi}(x_0), 
\end{eqnarray}
where 
\begin{eqnarray}
\label{eq2:6}
{\cal G}(k;x) \equiv e^{{\cal H}(k)x} = 
\left(
\begin{array}{cc}
            \cos kx & -k\sin kx \\
\frac{1}{k} \sin kx & \cos kx \\
\end{array}
\right) .
\end{eqnarray}
Because of the property $Tr {\cal H}(k)=0$, 
we have  ${\cal G}(k;x) \in SL(2,{\bf R})$, 
namely $\det {\cal G}(k;x) = 1$. 

We now consider a potential consisting of
three nearby $\delta$ functions 
located with equal distances $a$ which we assume to be small [7];  
\begin{eqnarray}
\label{eq2:8}
V(a;x) = v_1 \delta(x+a) + v_0 \delta(x) + v_1 \delta(x-a). 
\end{eqnarray}
The connection condition
between $x=(-a)_-$ and $x=a_+$ is given by 
\begin{eqnarray}
\label{eq2:9}
{\bf \Psi}(a_+) = {\cal V}_{\varepsilon;a}(k){\bf \Psi}((-a)_-)  
\end{eqnarray}
where
\begin{eqnarray}
\label{eq2:10}
{\cal V}_{\varepsilon;a}(k) = 
{\cal V}_{\delta}(v_1)
{\cal G}(k;a) 
{\cal V}_{\delta}(v_0)
{\cal G}(k;a) 
{\cal V}_{\delta}(v_1). 
\end{eqnarray}
For sufficiently small $a$ $\ll 1/k$, we can expand
Eq.(\ref{eq2:6}) as
\begin{eqnarray}
\label{eq2:11}
{\cal G}(k;a) \simeq 
\left(
\begin{array}{cc}
1 & -k^2a \\
a & 1 \\
\end{array}
\right)  
\end{eqnarray}
Inserting Eqs.(\ref{eq1:3}) and (\ref{eq2:11}) into 
Eq.(\ref{eq2:10}), 
we obtain 
\begin{eqnarray}
\left[{\cal V}_{\varepsilon;a}(k)\right]_{11} & \hspace{-1.5ex} = & 
\hspace{-1.5ex}\left[{\cal V}_{\varepsilon;a}(k)\right]_{22} \nonumber \\
\label{eq2:12a}
& \hspace{-1.5ex} = & 
\hspace{-1.5ex} 1+ ( v_0 + 2v_1 + v_0 v_1 a - k^2 a) a, \\ 
\label{eq2:12b}
\left[{\cal V}_{\varepsilon;a}(k)\right]_{12} & \hspace{-1.5ex} = & 
\hspace{-1.5ex} v_0 + 2 v_1 + 2 v_1 a (v_0 + v_1 - k^2 a) \nonumber \\
& & \hspace{-1.5ex} + v_0 v_1^2 a^2 - 2 k^2 a,  \\
\label{eq2:12c}
\left[{\cal V}_{\varepsilon;a}(k)\right]_{21} & \hspace{-1.5ex}= & 
\hspace{-1.5ex} 2a + v_0 a^2 . 
\end{eqnarray}
In the limit $a\rightarrow +0$ with $v_0$, $v_1$ constant, 
we simply obtain the connection condition for 
a $\delta$ potential of strength $v_0 + 2 v_1$. 
On the other hand, 
Eq.(\ref{eq2:12c}) tells that if 
the coupling strength $v_0$ varies as 
\begin{eqnarray}
\label{eq2:13}
v_0(a) \simeq \frac{u}{a^2} 
\end{eqnarray}
for small $a$, one obtains 
$ \displaystyle
\lim_{a\rightarrow +0} 
\left[{\cal V}_{\varepsilon;a}(k)\right]_{21} = u 
$. 
Furthermore, if we let
\begin{eqnarray}
\label{eq2:15}
v_1(a) \simeq \frac{2}{u}-\frac{1}{a}  ,
\end{eqnarray}
the diagonal elements (\ref{eq2:12a}) converge toward one
in the small $a$ limit. 
We obtain 
\begin{eqnarray}
\label{eq2:16}
\lim_{a\rightarrow +0} 
{\cal V}_{\varepsilon;a}(k) = 
{\cal V}_{\varepsilon} (u)    
\end{eqnarray}
with Eqs.(\ref{eq2:13}) and (\ref{eq2:15}). 
Thus our first objective is achieved.
Though both the strengths $v_0$, $v_1$ 
diverge in the small $a$ limit, 
$V(a;x)$ is nonsingular by construction.  
Note that 
$ \displaystyle
\lim_{a\rightarrow +0} 
\left[{\cal V}_{\varepsilon;a}(k) \right]_{12} = 0
$
is ensured because of  ${\cal V}_{\varepsilon;a}(k) \in SL(2,{\bf R})$. 

By using the $\varepsilon$ potential
and the $\delta$ potential in combination, 
one can realize the general connection condition 
\begin{eqnarray}
\label{eq2:18}
{\cal V}=
\left(
\begin{array}{cc}
t & v \\
u & s 
\end{array}
\right), 
\end{eqnarray}
where $t s - u v = 1$, 
$t, v, u, s \in {\bf R}$. 
The following matrix decompositions serve for our purpose;   
\begin{eqnarray}
\label{eq2:19}
\left( 
\begin{array}{cc}
t & v \\
u & s 
\end{array}
\right)
= 
\left\{
\begin{array}{ll}
{\cal V}_{\delta}(\frac{t-1}{u})
{\cal V}_{\varepsilon}(u)
{\cal V}_{\delta}(\frac{s-1}{u}) &
\mbox{for} \hspace{1ex} u \neq 0, \\[2ex]
{\cal V}_{\varepsilon}(\frac{s-1}{v})
{\cal V}_{\delta}(v)
{\cal V}_{\varepsilon}(\frac{t-1}{v}) & 
\mbox{for} \hspace{1ex} v \neq 0. 
\end{array} 
\right. \nonumber
\end{eqnarray}
\vspace*{-5ex}
\begin{eqnarray}
\label{eq2:20}
\end{eqnarray}
Each matrix on RHS takes a form of  
either $\delta$ or $\varepsilon$ potential. 
This implies that 
in case of $u \neq 0$, for instance, 
one can realize a general connection condition 
in the limit 
{\small
\begin{eqnarray}
\label{eq2:21}
V(x) = \lim_{b\rightarrow +0} 
\left(
\frac{s-1}{u}
\delta (x+b)+
u \varepsilon(x)+
\frac{t-1}{u}
\delta (x-b)
\right). \nonumber 
\end{eqnarray}
}
\vspace*{-3ex}
\begin{eqnarray}
\label{eq2:22}
\end{eqnarray}
In case of $v = u =0$, 
one can use the decomposition 
\begin{eqnarray}
\label{eq2:23}
\left(
\begin{array}{cc}
\pm |t| & 0 \\
     0  & \pm |s| 
\end{array}
\right)
& \hspace{-1ex} = & \hspace{-1ex}  
{\cal V}_{\delta}(\rho)
{\cal V}_{\varepsilon}(-\frac{1}{\rho})
{\cal V}_{\delta}(\rho) \nonumber \\
& \hspace{-1ex} \times & \hspace{-1ex} 
{\cal V}_{\delta}(\mp\frac{1}{\rho})
{\cal V}_{\varepsilon}(\pm\rho)
{\cal V}_{\delta}(\mp\frac{1}{\rho})
\end{eqnarray}
with $\rho=\sqrt{|t|}$, which is also written 
in terms of the connection matrices for the $\delta$ and  
$\varepsilon$ potentials. 

\bigskip
\bigskip
\noindent{\large\bf III. 
Scattering Properties and Delta-Epsilon Duality}

\medskip
In this section, 
we briefly discuss the scattering properties 
of the $\varepsilon$ potential and point out a 
dual role between the $\delta$ and $\varepsilon$ potentials. 

We start by putting a general contact interaction at the origin 
on $x$-axis. 
Incident and outgoing waves can be written as 
\begin{eqnarray}
\label{eq3:1}
\varphi_{in}(x) & = & A(k) e^{ikx} + B(k) e^{-ikx} 
\ \ \ (x < 0), \\
\label{eq3:2}
\varphi_{out}(x) &  = & e^{ikx}
\ \ \ \ \ \ \ \ \ \ \ \ \
\ \ \ \ \ \ \ \ \ \ \ \ \
 (x > 0).  
\end{eqnarray}
Here we assume the incident wave comes from minus infinity 
and as a result, $\varphi_{out}$ has only an outgoing component,   
whose amplitude is normalized to one. 
The connection condition in Eq.(\ref{eq1:2}) is written as  
\begin{eqnarray}
\label{eq3:3}
\left( \! \!
\begin{array}{c}
ik \\
1 
\end{array}
\! \! \right)
=
{\cal V} 
\left(
\begin{array}{cc}
ik & -ik \\
1  &   1 
\end{array}
\right)
\left( \! \!
\begin{array}{c}
A(k) \\
B(k) 
\end{array}
\! \! \right),  
\end{eqnarray}
leading to 
\begin{eqnarray}
\label{eq3:4}
\left( \! \!
\begin{array}{c}
A(k) \\
B(k) 
\end{array}
\! \! \right)
= 
\frac{1}{2ik}
\left(
\begin{array}{cc}
 1  &  ik \\
-1  &  ik 
\end{array}
\right)
{\cal V}^{-1}
\left( \! \!
\begin{array}{c}
ik \\
1 
\end{array}
\! \! \right). 
\end{eqnarray}
The transmission and reflection coefficients are calculated with
\begin{eqnarray}
\label{eq3:5}
T(k)=\left| \frac{1}{A(k)} \right|^2, \hspace{5ex}
R(k)=\left| \frac{B(k)}{A(k)} \right|^2, 
\end{eqnarray}
respectively. 
In case of ${\cal V}={\cal V}_{\delta}(v)$, 
we obtain the well-known results  
\begin{eqnarray}
\label{eq3:6}
T_{\delta}(k) = \frac{k^2}{k^2+(v/2)^2}, \hspace{2ex}
R_{\delta}(k) = \frac{(v/2)^2}{k^2+(v/2)^2}. 
\end{eqnarray}
For ${\cal V}={\cal V}_{\varepsilon}(u)$, we obtain 
\begin{eqnarray}
\label{eq3:7}
T_{\varepsilon}(k) = \frac{(2/u)^2}{k^2+(2/u)^2}, \hspace{2ex}
R_{\varepsilon}(k) = \frac{k^2}{k^2+(2/u)^2} .
\end{eqnarray}
One can observe that if $u=v$,  
$T_{\delta}(k)=T_{\varepsilon}(1/k)$ and 
$R_{\delta}(k)=R_{\varepsilon}(1/k)$ are satisfied. 
This implies that the low (resp. high) energy dynamics of 
$\varepsilon$ potential is described by the high (resp. low) 
energy dynamics of $\delta$ potential. 

The dual role of $\delta$ and $\varepsilon$ potentials becomes
more manifest 
when we consider the scattering of two {\it identical}
particles.  We now regard the variable $x$ as the relative 
coordinate of two identical particles whose statistics 
is either fermionic or bosonic.  The incoming and outgoing waves
are now related by the exchange symmetry.  We assume the form
\begin{eqnarray}
\label{eq3:8}
\varphi_{in}(x) & = &  e^{ikx} + C(k) e^{-ikx}
\ \ \ \ \ \ \ \ \ \ \ \ \
(x < 0) ,\\
\label{eq3:9}
\varphi_{out}(x) &  = & \pm e^{-ikx} \pm C(k) e^{ikx}
\ \ \ \ \ \ \ \ \ \ \
(x > 0) .\\
\end{eqnarray}
where the composite signs take $+$ for bosons and $-$ for fermions.
Important fact to note is that the symmetry (resp. anisymmetry) of
$\varphi(x)$ leads to the antisymmetry (resp. symmetry) of it's 
derivative $\varphi'(x)$.
The coefficient $C(k)$ becomes the scattering matrix.
The connection condition in Eq.(\ref{eq1:2}) now reads
\begin{eqnarray}
\label{eq3:10}
\left[
  \left(
    \begin{array}{cc}
     -ik & ik \\
      1  &   1 
     \end{array}
    \right)
  \mp
    {\cal V} 
   \left(
     \begin{array}{cc}
       ik & -ik \\
       1  &   1 
      \end{array}
   \right)
\right]
\left( \! \!
\begin{array}{c}
1 \\
C(k) 
\end{array}
\! \! \right)
= 0 .
\end{eqnarray}
We first consider the case for $\delta$ potential 
${\cal V}$ $ = {\cal V}_\delta(v)$.
One obtains 
\begin{eqnarray}
\label{eq3:11}
C_\delta(k) &=& 1   \ \ \ \ \ \ \ \ \ \ \ \ \ \ 
  {\rm for \ fermions}, \\
C_\delta(k) &=& {{2ik + v} \over {2ik - v}} \ \ \ \ \ 
  {\rm for \ bosons}.
\end{eqnarray}
The first equation means that the $\delta$ function is
inoperative as the two-body interaction between identical
bosons, which is of course a trivial fact.
Next we consider the case of $\varepsilon$ potential 
${\cal V}$ $ = {\cal V}_\varepsilon(u)$. 
One has
\begin{eqnarray}
\label{eq3:12}
C_\varepsilon(k) &=& {{2ik + 4/u} \over {2ik - 4/u}} \ \ \ 
  {\rm for \ fermions}, \\
C_\varepsilon(k) &=& 1 \ \ \ \ \ \ \ \ \ \ \ \ \ \
  {\rm for \ bosons}.
\end{eqnarray}
One finds that the role of fermion and boson cases are exchanged:
The $\varepsilon$ potential as the two-body interaction  has 
no effect on identical bosons,
but {\it does} have an effect on the fermions.
Moreover, the scattering amplitude of fermions with
${\cal V}_\varepsilon(u)$ is exactly the same as 
that of bosons with ${\cal V}_\delta(v)$ if the two
coupling constants are related as
\begin{eqnarray}
\label{eq3:13}
vu = 4.
\end{eqnarray}
Therefore, a two-fermion system with $\varepsilon$ potential 
is dual to a two-boson system with $\delta$ potential.  
As expected, a natural generalization to 
$N$-particle systems exists [8].

\bigskip
\bigskip
%
%
\noindent{\large\bf IV. Conclusion}

\medskip
From the construction Eq. (\ref{eq2:19}),
we can easily see the reason why 
the generalized contact interaction
in one-dimension with time reversal symmetry is 
of three parameter family: There are two kinds of
``elementary'' contact interactions -- 
$\delta$ and $\varepsilon$ -- and one obtains 
different physics by different spatial ordering of
$\delta$ and $\varepsilon$ even in the zero-range limit.
This construction should be also useful in the actual design
of the possible device realizing the generalized contact 
interaction in mesoscopic scale.

It is clear 
from the arguments in the last section 
that the $\varepsilon$ interaction
is a generic short-range limit
of arbitrary two-body interaction between like fermions,
just as $\delta$ interaction being the generic zero-range
force between like bosons.
As such, the $\varepsilon$ interaction, which has been 
a rather mysterious object, should be viewed
as an integral component
of one-dimensional many-body quantum mechanics.

\medskip
We are grateful to Prof. Y. Okada for useful discussions.
This work has been supported in part by the Grant-in-Aid 
(No. 10640396) by the Japanese Ministry of Education.

\bigskip
\bigskip
\noindent{\large\bf References}

\smallskip
\noindent{[1]}
{
S. Albeverio, F. Gesztesy, R. H{\o}egh-Krohn and H. Holden, 
``Solvable Models in Quantum Mechanics,'' 
Springer-Verlag, New York, 1988.
}

\noindent{[2]}
{
P. Seba,
Czech. J. Phys., B36 (1986) 667.
}

\noindent{[3]}
{
P. Seba,
Rep. Math. Phys. 24 (1986) 111.
}

\noindent{[4]}
{
M. Careau,
J. Phys. A26 (1993) 427.
}

\noindent{[5]}
{
P.R. Chernoff and R. Hughes,
J. Funct. Anal. 111 (1993) 92.
}

\noindent{[6]}
{
J.E. Avron, P. Exner, Y. Last,
Phys. Rev. Lett. 72 (1994) 896.
}

\noindent{[7]}
{
T. Cheon and T. Shigehara, 
Phys. Lett. A243 (1998) 111.
}

\noindent{[8]}
{
T. Cheon and T. Shigehara, 
LANL preprint quant-ph/9806041.  
}

%
%

\end{document}